\documentclass[onecolumn,aps,pra,preprintnumbers,amsmath,amssymb]{revtex4}
\usepackage{natbib}
\usepackage{graphicx}
\usepackage{bm}
\usepackage{color}

\usepackage{amsmath}
\usepackage{amsfonts}
\usepackage{amssymb}
\usepackage{hyperref}
\usepackage{makeidx}
\usepackage{graphicx}
\usepackage{physics}

\begin{document}
	
	\title{Time evolution of quantum correlations in presence of state dependent bath}
	\author{Mehboob Rashid$^1$}
	\author{Muzaffar Qadir Lone$^2$}
	\author{Prince A Ganai$^1$}
	\affiliation{$^1$Department of Physics, National Institute of Technology, Srinagar-190006 India.\\
		$^2$ Department of Physics, University of Kashmir, Srinagar-190006 India}
	
	\begin{abstract}
		The emerging quantum technologies heavily rely on the understanding of dynamics in open quantum systems. In the Born approximation, the initial system-bath correlations are often neglected which can be violated in the strong coupling regimes and quantum state preparation. In order to understand the influence of initial system-bath correlations, we study the extent to which these initial  correlations and the distance of separation between qubits influence the  dynamics of quantum entanglement and coherence. It is shown that at low temperatures, the initial correlations have no role to play while at high temperatures, these correlations strongly influence the dynamics. Furthermore, we have shown that distance of separation between the qubits in presence of collective bath helps to maintain entanglement and coherence at long times.

	\end{abstract}
	\maketitle
	\section{Introduction}
	Quantum correlations (entanglement) \cite{1} form a key concept towards the foundational understanding of quantum mechanics, and has been realized as a   precious resource for various tasks that are impossible in classical domian like quantum teleportation, cryptograpgy, and various other processings\cite{2,3,4,5}.The deeper insights into the dynamics of  quantum correlations is of great importance, since the real quantum system always interacts with its bath that leads to decoherence of quantum superpositions and entanglement degradation \cite{6,7,8,9,10}. This system-bath dynamics is categorized either as Markovian or non-Markovian \cite{11,12,13}. In a Markov process, the system loses 	information to environment irreversibly. However, many quantum systems posses a pronounced
	non-Markovian behavior in which there is flow of information back to system,  signifying the presence of quantum memory effects \cite{14,15,16,17,18,19,20,21,22}. The physical realization and
	control of dynamical processes in open quantum systems plays a decisive role, for example, in recent proposals for the generation of entangled states \cite{23,24}, for schemes of  dissipative quantum computation \cite{25,26}, for the design of quantum memories \cite{27} and for the enhancement of the efficiency in quantum metrology \cite{28}.  Moreover, in the regime 	where non-Markovian effects are important, the presence of system-bath correlations invalidates the initial state in which the system and the bath are independent. Especially in the experimentally relevant case where the qubit system is excited out of equilibrium  and the subsequent dynamics is probed, a proper treatment of the initial state is crucial.

	A complete understanding of the dynamics of entanglement \cite{8,9,10} relies on available measure that can reflect the time variation of the system of interest. Here, we utilize Concurrence \cite{29,30} as the entanglement measure to understand the underlying dynamics in presence of initial system-bath correlations.
	In addition to the quantum entanglement, coherence  has been proposed as an alternative resouce for quantum information processings \cite{31,32,33,34,35}.It has attracted much attention over the past decade both in theory and experiment \cite{36,37,38,39}.  It has been shown  that the coherence exists in photosynthetic complexes \cite{40,41,42}, therefore can play an important role in explaining the high efficiency of the these complexes, which in turn have technological benifits. There exist several quantifiers for coherence \cite{32} in a given quantum system like $l_1$-norm, relative entropy etc. In this work, we utilize $l_1-$norm which is in simple terms represents the sum of off-diagonal terms of a density matrix.	Due to its important role in quantum mechanics\cite{43,44}, quantum information \cite{45,46}, and quantum biology \cite{47,48,49,50}, the behavior of coherence during the evolution of a system is necessary 	to be investigated.

	There has been tremendous amount of work towards understanding the non-Markovian behaviour in single and many qubits coupled to a bosonic bath\cite{51,52,53,54,55,56,57}. In most of these cases, Born approximation is used which means  that  the joint initial state of the system and bath 
	are assumed to be uncorrelated. However this assumption is often violated when there exist strong interaction between system and the bath \cite{21,58,59,60,61,62,63,64,65,66}. In particular, quantum state preparation  can lead to strong system-bath correlations thus effecting the subsequent dynamics. In this regard, various works have critcally  examined the initial system-bath correlations in spin-Boson model\cite{67}, superconducting qubits \cite{68}, quantum dots \cite{69} etc. For example, Mozorov et. al. \cite{61,67}, has considered an exactly solvable dephasing model of a single qubit interacting a bosonic in presence of initial system-bath correlations, while in \cite{70} has considered Jaynes-Cunnings model and \cite{71} has studied  in detail dynamics  in presence of spin bath. These models however, consider either the single qubit or many coupled with individual baths. Furthermore, the distance of separation between the qubits in presence of initial correlations is  not taken into account in these works.  Our main objective in this work is to study the effect of distance dependent interaction between qubits and the collective bath in presence of initial system-bath correlations. Such kind of settings can be obtained using cold atom impurities immersed in a Bose-Einestien condensate (BEC) \cite{72}. The positioning of immersing of cold atoms in the BEC would yeild a distance dependent interaction with non-trivial bath spectral density. Thus the present work would be mainly  important for quantum information processings using cold atoms. 	
	
	This paper is organised in the following way. In section II, we introduce  the  model of qubits interacting with a collective bath in presence of initial system-bath correlations and calculate the time evolution of two qubits density matrix. In section III, we discuss the dynamics of entanglement given by concurrence and coherence. Finally we conclude in section IV.
	
	\section{Model Calculations}
	In this section, we introduce our model of  qubits interacting with a bosonic bath. In spin representation, the total Hamiltonian of the model is written as ( $\hbar=1$). 
	\begin{eqnarray}
		\label{1}
		H &=& H_S + H_B + H_{int} \nonumber \\
		&=&\frac{\omega_0}{2}\sum_i\sigma_{i}^z + \sum_{k}\omega_{k}b_{k}^{\dagger}b_k + \sum_{ik}\sigma_{i}^z(g_{k}e^{-i\vec{k}.\vec{r_i}}b_k + h.c.) \nonumber \\	
	\end{eqnarray}
	where $\omega_0$ is the energy splitting of the qubit, $\sigma_{z}$ is the $z$-Pauli matrix, $\vec{r_i}$ is the position vector of $ith$-qubit, and h.c. means Hermitian conjugate.  We assume distance of separation between two qubits to $\vec{L}=\vec{r}_1-\vec{r}_2$. $g_k$ is the system-bath coupling. This kind of model can be realized in ultracold setting by immersing an ultracold gas trapped in an optical lattice in Bose Einestien condensate \cite{72,73}. The low lying exciations of BEC i.e. Bogoliubov phonons will act as the bosonic bath. We assume linear dispersion for bath modes (phonon or photon type). In BEC setting it will correspond to phonon like excitations in long wavelength domain instead of particle like excitations at large momentum. 
	
	In this work, we will assume a particular type of initial state of the bath and system as a case study to dynamics of quantum correlations.  Before $ t = 0 $, we initially start in the state for the combined system and bath that are thermal states of the  whole system 
	\begin{align}
		\rho  = \frac{e^{-\beta H }}{Z}
		\label{thermalbath2}
	\end{align}
	where $ Z $ is the partion function and $\beta =\frac{1}{T}$. 
	This should be compared to usual situtaion  where the thermal state is considered  only with respect to the bath states i.e. $\rho_B= \frac{e^{-\beta H_B }}{Z_B}$. This type of state could arise in a situation where the measurement apparatus is prepared in the vicinity of the system prior to the state preparation.  Since the system is rather small, the time taken to reach thermal equilibrium can be rather short, and the state (\ref{thermalbath2}) is attained quickly before the state preparation is performed. 
	
	Now we would like to prepare the state of the system in a manner such that  a projective measurement is made on the system. For the system the projection operators are
	\begin{align}
		P_S  = \{ | \psi \rangle \langle \psi |, | \psi_\perp \rangle \langle \psi_\perp | \}
	\end{align}
	where $ | \psi \rangle $ is the  initial state of the system  and  $ | \psi_\perp  \rangle $ is the state that is orthogonal to this state.  Now we postselect on the state  $ | \psi \rangle $ such that the system is ensured to have the state 
	\begin{eqnarray}
		\rho_S (0) = |\psi\rangle \langle \psi|  .
	\end{eqnarray}
	and the bath states then take a form
	\begin{align}
		\rho_{B}^\psi = \frac{\bra{\psi}\exp(-\beta H)\ket{\psi}}{Tr_{B}\bra{\psi}\exp(-\beta H)\ket{\psi}}
	\end{align}
	where there is a dependence on the state of the system because the original state (\ref{thermalbath2}) were thermal states in the space of the system and bath.  The initial state of the whole system is then the product state
	\begin{align}
		\rho(0) = \rho_S (0) \otimes  \rho_B^{\psi} (0) .
	\end{align}
	The primary difference to uncorrelated case is that the bath state according to this preparation depends in a non-trivial way on the system state.  We assume the initial state of the system to be a general two qubit state $|\psi\rangle= a|00\rangle + b|01\rangle + c|10\rangle + d|11\rangle$ with $|a|^2 + |b|^2+|c|^2+|d|^2=1$. For this state, we have (for detailed calculations see appendix):
	\begin{eqnarray}
		\rho_B^{\psi} (0)=\frac{ 1}{Z}\big[ |a|^2 e^{-\beta\omega_0}e^{-\beta H_{B1}^{+}} +|b|^2 e^{-\beta H_{B1}^{-}}+|c|^2 e^{-\beta H_{B2}^{+} }+ |d|^2 e^{\beta\omega_0} e^{-\beta H_{B2}^{-} }\big]
	\end{eqnarray}
	 Next, our main interest is to calculate the reduced density matrix of the system by tracing out degrees of freedom of the bath:
	
\begin{eqnarray}
	\rho_S(t)= {\rm Tr}_B[  U(t)\rho(0)U(t)^{\dagger} ]
\end{eqnarray}
where $U(t)=T~e^{-i\int_0^t dt^{\prime} H_I(t^{\prime})}$ is the time evolution operator and $H_I(t)$ is the interaction
Hamiltonian in interactin picture. 
We write 
$U(t) = e^{i\phi(t)}e^{\Sigma_{i}\sigma_{i}^{z}\hat\Lambda_i(t)}$ where $\phi(t)$
is a function of time only and $\hat\Lambda_i(t) = \Sigma_{k}(\alpha_{ik}(t)b_k - \alpha_{ik}^{*}(t)b_k^{\dagger})$ with $\alpha_{ik}(t) = g_{ik}\frac{(1-e^{-i\omega_{k}t})}{\omega_{k}}$ and $\phi(t)$ is a function of time only. Therefore, we can write
\begin{eqnarray}
	\rho_S(t) &=& {\rm Tr}_B[U(t) \rho(0) U(t)^{\dagger}]\\
	&=& {\rm Tr}_B[ e^{\Sigma_{i}\sigma_{i}^{z}\hat\Lambda_i(t)} |\psi\rangle \langle \psi| \otimes 
	\rho_B^{\psi}  e^{-\Sigma_{i}\sigma_{i}^{z}\hat\Lambda_i(t)}  ]
\end{eqnarray}
after a cumbersome calculations (see appendix ), we get
\begin{eqnarray}\label{densitymatrix}
	\rho_{s}(t)=\begin{pmatrix}
		|a|^2 & ab^* \varphi(t) & ac^*\zeta(t)& ad^* \kappa(t)\\
		ba^* \varphi^*(t) & |b|^2 & bc^*\bar{\kappa}(t) &bd^*\bar{\zeta}(t)\\
		ca^*\zeta^*(t)&cb^*\bar{\kappa}^*(t)&|c|^2&cd^*\bar{\varphi}(t)\\
		da^* \kappa^*(t)&db^*\bar{\zeta}^*(t)&dc^* \bar{\varphi}^*(t)& |d|^2
	\end{pmatrix}
\end{eqnarray} 

where the different functions are written explicitly in appendix. In the next section we examine the entanglement and coherence in various approximations.

\section{Dynamics of Entanglement and Coherence}

\subsection{Concurrence}
In this section, we  analyze quantum entanglement measured by concurrence in a two qubit system considered in this work. For a density matrix $\rho$, the concurrence is defined as 
\begin{eqnarray}
	\mathcal{C}= {\rm max} \{ 0, \sqrt{\lambda_1}- \sqrt{\lambda_2}- \sqrt{\lambda_3}- \sqrt{\lambda_4}\},
\end{eqnarray}
where $\lambda_i,i=1,2,3,4$ are the eigen values of the matrix $\rho\tilde{\rho}$ taken in descending order and  $\tilde{\rho}=(\sigma^y_1\otimes \sigma^y_2) \rho^{\star}( \sigma^y_1\otimes \sigma^y_2)$ is the time reversal density matrix. $\rho^{\star}$ is the conjugation obtained in standard basis. For an unentangled state $\mathcal{C}=0$ while for maximally entangled state $\mathcal{C}=1$. In order to simplify our calculations,
we assume $a=\sqrt{p}$, $d=\sqrt{1-p}$ and $b=0=c$ so that  our initial state is  the class of states $\psi=\sqrt{p}|00\rangle + \sqrt{1-p}|11\rangle $, with $0\le p\le 1$.
The same analysis holds for other types of states as well.  For this state, time evolved concurrence can be written as
\begin{eqnarray}
	\mathcal{C}(t)= 2{\rm max} \{ 0, C(t)\},
\end{eqnarray}
with 
\begin{eqnarray}
	C(t)&=& \sqrt{\cos^2\Phi + \sin^2\Phi \tanh^2(\beta\omega_0)} e^{-\gamma(t)} C(0)
	\label{con}
\end{eqnarray}
where
\begin{eqnarray}
	\Phi(t)&=& 8 \sum_k \frac{|g_k|^2}{\omega_k^2} \sin \omega_k t \Big[1+\cos (\vec{k}.\vec{L})\Big] \nonumber \\
	\gamma (t)&=& 4\sum_k |g_k|^2 \Big[1+\cos (\vec{k}.\vec{L})\Big] \frac{1-\cos \omega_k t}{\omega_k^2} \coth \frac{\beta \omega_k}{2} \nonumber \\
\end{eqnarray}
We rewrite the concurrence $C(t)=e^{-\Gamma(t)} C(0)$, with $C(0)=2\sqrt{p(1-p)}$ as the initial entanglement of the system. Thus unentangled states $p=0$ and $p=1$ remain always unetangled. This is due to the dephasing nature of the interaction. Thus all values of $0<p<1$ show the same type of behavour. Therefore, without loss any generality we take $p=1/2$. Also, $ \Gamma(t)= \gamma_0(t)+ \gamma (t)$ with
\begin{eqnarray}
 \gamma_0(t) =-\frac{1}{2} \log [ \cos^2\Phi + \sin^2\Phi \tanh^2(\beta\omega_0) ]
\end{eqnarray}
 as the decoherence function due to initial correlations and $\gamma (t)$ is the standard decoherence function which is present even in absence of initial system-bath correlations.

\subsection{Coherence}
Quantum coherence directly stems from the superposition principle that enables it to show quantum interference phenomena. There has been several proposals to quantify coherence in a given quantum system. However, we take $l_1$-norm quantification of coherence in the present work. It is an intuitive measure related to off-diagonal terms of the density matrix. It is defined as 
\begin{eqnarray}
	\mathcal{N}= \sum_{i\ne j}||\rho_{ij}-\mathcal{I}_{ij} ||
\end{eqnarray}
where the optimization is to be carried out over all possible incoherent states $\mathcal{I}$. After the optimization, we get the following expression for the coherence in the standard basis
\begin{eqnarray}
	\mathcal{N} = \sum_{i\ne j}|\langle i |\rho |j\rangle|.
\end{eqnarray}
This simply represents the sum of off-diagonal elements of the density matrix under consideration and therefore  captures the notion of interference in a quantum state. For the states under consideration we have
\begin{eqnarray}
	\mathcal{N}=2 \sqrt{\cos^2\Phi + \sin^2\Phi \tanh^2(\beta\omega_0)} e^{-\gamma(t)}= 2 C(t).
\end{eqnarray}
This is true for other types of  states($ \sqrt{p}|01\rangle + \sqrt{1-p}|10\rangle$) as well. Thus, pairwise entanglemet given by concurrence is half to that of the coherence given by $l_1$ norm. Thus it is sufficient to analyze the concurrence $C(t)$. Furthermore, this provides an explicit example where entanglement can be measured by coherence\cite{34,38}.

\subsection{Decay of Concurrence}

Next, we analyze the decoherenc functions $\Phi(t)$ and $\gamma (t)$ using various approximations. Before, evaluating the sum in these equations, we realize that $\gamma(t)$ becomes zero for certian modes of the bath. For $\vec{k}.\vec{L}= (2n+1) \pi, n=0, 1,..$, we have $\cos \vec{k}.\vec{L}=-1 $, which makes $\gamma (t)=0$ and  $\Phi(t)=0$. It implies there exist certian bath  modes for $\theta = \cos^{-1}\Big( \frac{(2n+1)\pi}{kL}\Big)$  that do not lead to the decay of entanglement (or coherence), $\theta$ is the angle  between  $\vec{k}$ and   $\vec{L}$. In other words, all bath modes do not couple to the qubits which can lead to decoherence.  This can be intuitively understood from the fact that these bath modes do not  resolve  the separation of  qubits, hence suppresses the decay of correlations. Next, we define the bath spectral density as 
\begin{eqnarray}
	J(\omega)= \sum_{k}|g_{1k}+g_{2k}|^2 \delta(\omega-\omega_k) 
\end{eqnarray}
where $g_{ik}=g_ke^{-i\vec{k}.\vec{r}_i}$. This spectral density has a very complicated form and in general depends on the dimensionality $d$ of the bath.  This structure of the spectral density requires the bath to resolve the distance between the qubits, hence the  factor of $(1+\cos \vec{k}.\vec{L})$. Next, assume that the form of  $|g(\omega)|^2=g_0\frac{\omega}{\Omega^2} e^{-\omega^2/\Omega^2}$, \cite{74} where $g_0$ is the intrinsinc  coupling between system and the bath, $\Omega$ is the cut-off frequency of the bath.  Therefore, we write
\begin{eqnarray}
	\label{gamma2}
	\gamma(t) &=&16 \int \frac{d^3k}{(2\pi)^3} |g_k|^2 \cos^2(\vec{k}.\vec{L}) \frac{\sin^2 \frac{\omega_k t}{2}}{\omega_k^2} \coth\frac{\beta \omega_k}{2} \nonumber \\
&=& \frac{4 g_0}{ \Omega^2 \pi^2 v^3} \int_0^{\infty} d\omega \omega e^{-\frac{\omega^2}{\Omega^2}}\Big(1+\frac{\sin \omega s}{\omega s}\Big) \sin^2 \frac{\omega_k t}{2}\coth\frac{\beta \omega}{2} 
\end{eqnarray}
and 
\begin{eqnarray}
	\label{phi2}
	\Phi(t)&=&  8\int \frac{d^3k}{(2\pi)^3} |g_k|^2 \cos^2(\vec{k}.\vec{L}) \frac{\sin \omega_k t}{\omega_k^2}  \nonumber \\
	&=& \frac{4g_0}{\pi^2 v^3\Omega^2} \int_0^{\infty} d\omega \omega e^{-\frac{\omega^2}{\Omega^2}}\Big(1+\frac{\sin \omega s}{\omega s}\Big) \sin \omega t 
\end{eqnarray}
where $s=\frac{L}{v}$ is the time scale provided by the interactions mediated by the bath modes between qubits separated by the distance $L$ and $v$ is the velocity of  bath modes. In a typical experimental setup \cite{75} for cold atoms, we can vary distance between two qubits from  $L=100$nm to $10 \mu $m with speed of sound $v=350ms^{-1}$, which yeilds $s=0.1$ns to $10$ns.
Also, we have different energy scales arising in our model. The highest energy scale is given by the cutt off frequency $\Omega$ which provides the relaxation time scale for the bath $\tau_B \sim \frac{1}{\Omega}$; the energy scale $\omega_0$ provides a natural time scale for the relaxation of qubits $\tau_s\sim \frac{1}{\omega_0}$. Now we can parametrize the above equations \ref{gamma2} and \ref{phi2} in the following way: $\omega \to \frac{\omega}{\Omega},~t \to \Omega t, s\to \Omega s$ and the temperature is measured with respect to $\Omega$: $ \beta \to \beta \Omega$. For notational convienence,  we take $\Omega=1$ without loss of generality.

\begin{figure}[t]
	\includegraphics[width=3.5cm,height=3cm]{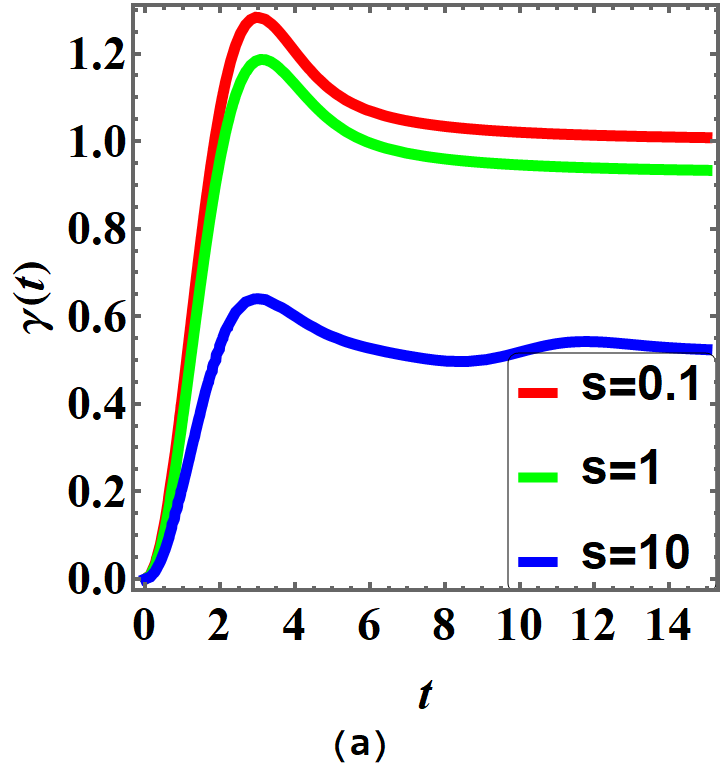} \hspace{0.5cm}
	\includegraphics[width=3.5cm,height=3cm]{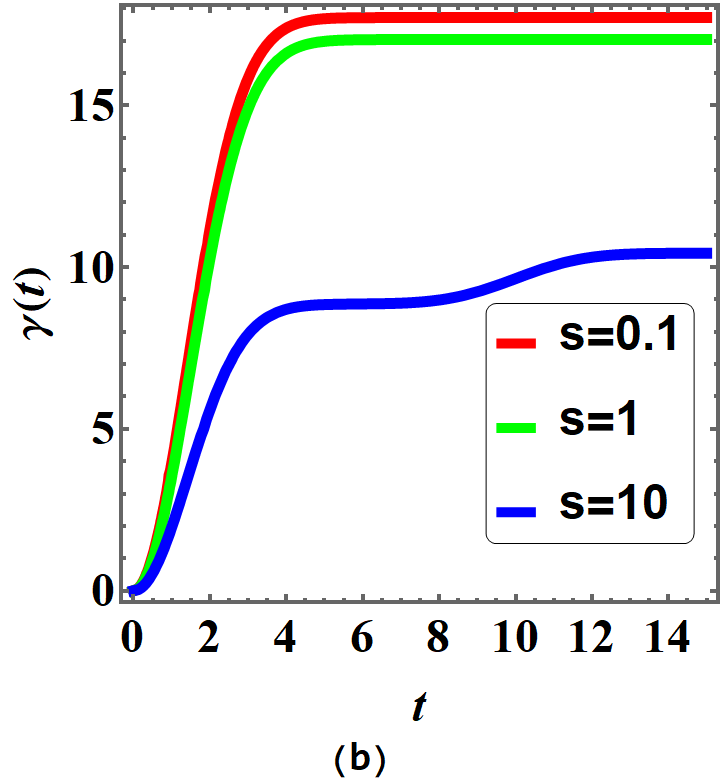}\\
	\includegraphics[width=3.5cm,height=3cm]{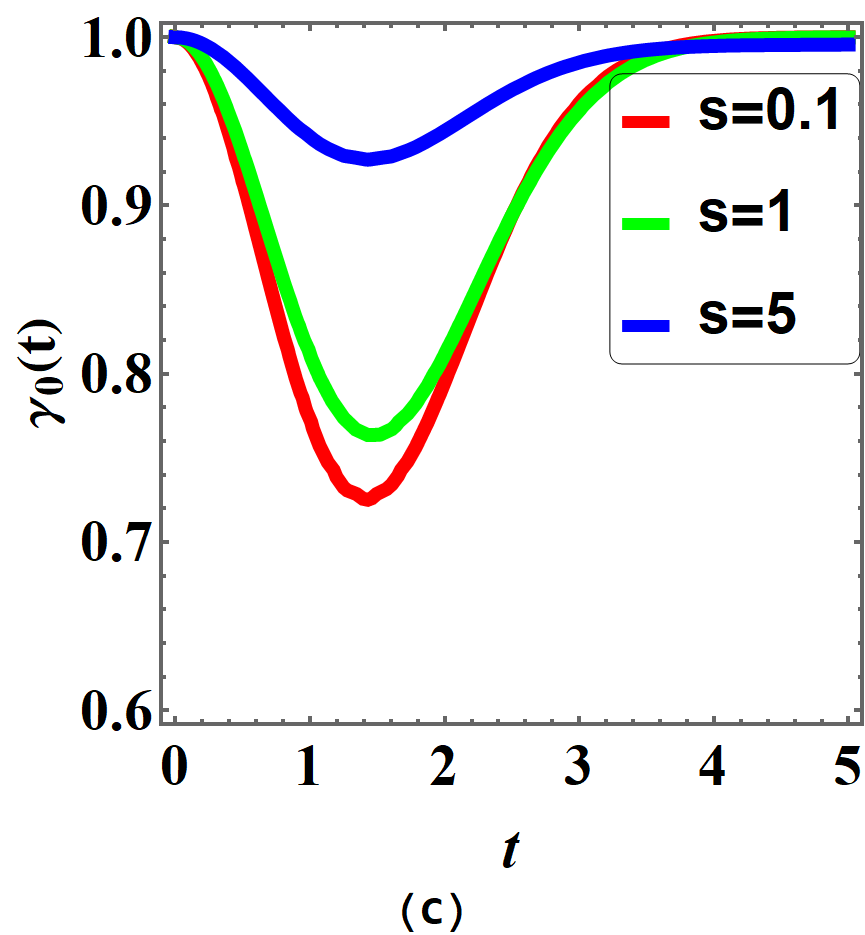}
	\hspace{0.5cm} 	  
\includegraphics[width=3.5cm,height=3cm]{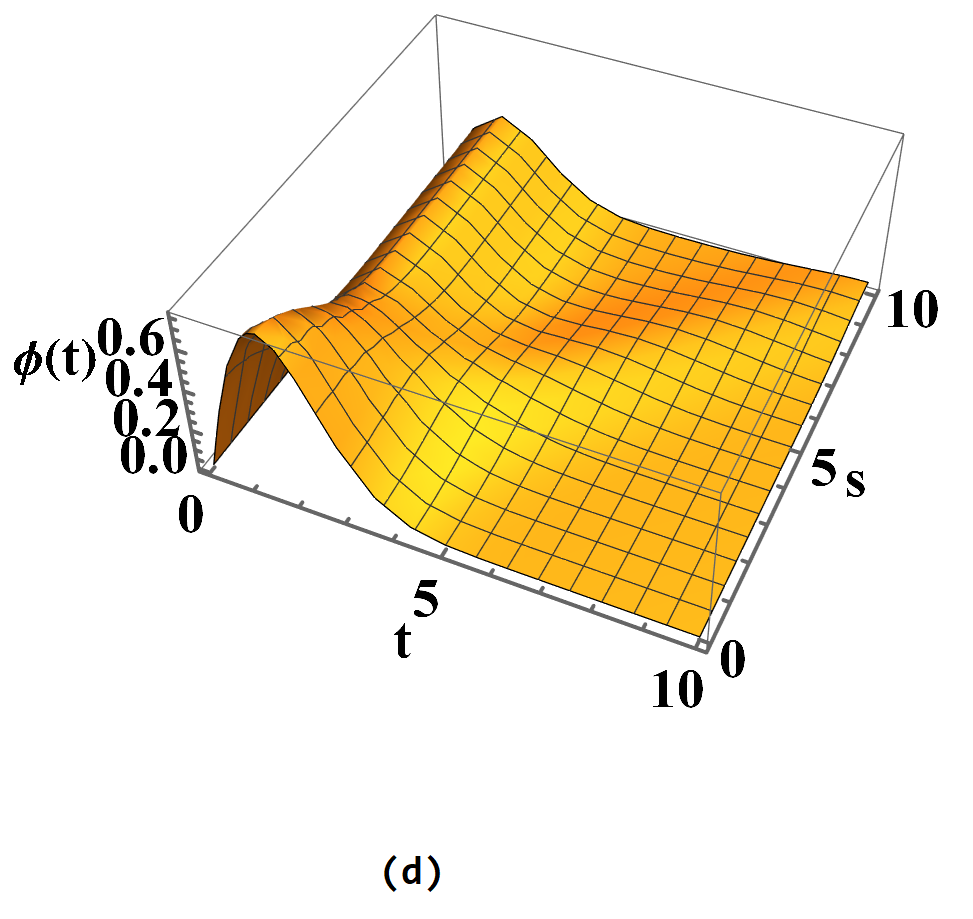}
	\caption{Variation of decoherence functions $\gamma(t)$, $\gamma_0(t)$ and depahsing function $\Phi(t)$ for different values of $s$: In (a) and (b) we have $\gamma(t)$ vs $t$ in units of $\Omega$ for low temperature  $\beta \Omega >>1$ and high temperature  $\beta\Omega <<1$ cases repectively. In (c) we have $\gamma_0(t)$ vs $t$ for  $\beta\Omega <<1$, (d)  $\Phi(t)$ does not have any  temperature dependence and varies non-monotonically with qubit separation.  In the long time limit, in all cases, $\gamma(t)$ saturates at some particular value, which suggest suprresion of decoherence due to long wavelength modes of the bath. }
	\label{fig1}
\end{figure}

	\begin{figure}[t]
	\includegraphics[width=3.5cm,height=3cm]{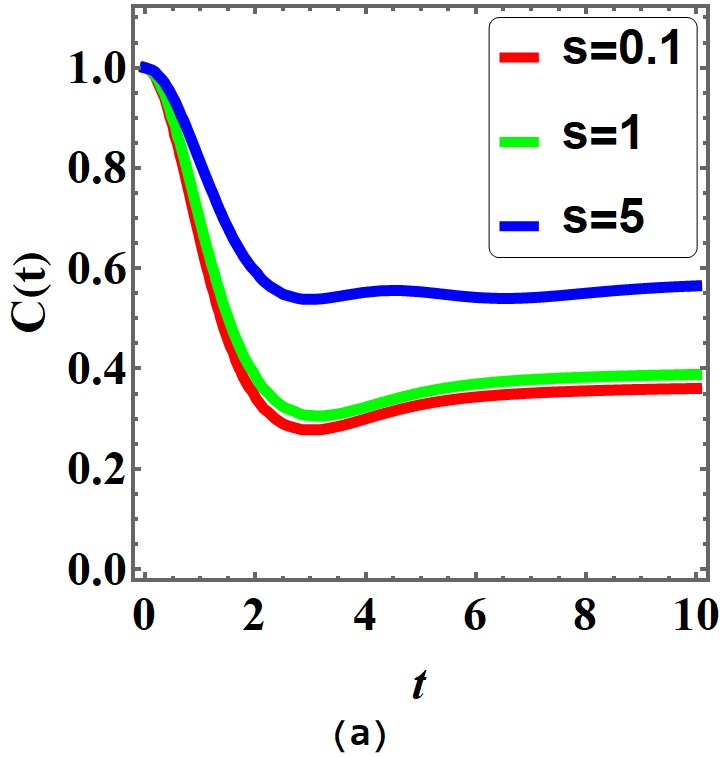} \hspace{0.5cm}
	\includegraphics[width=3.5cm,height=3cm]{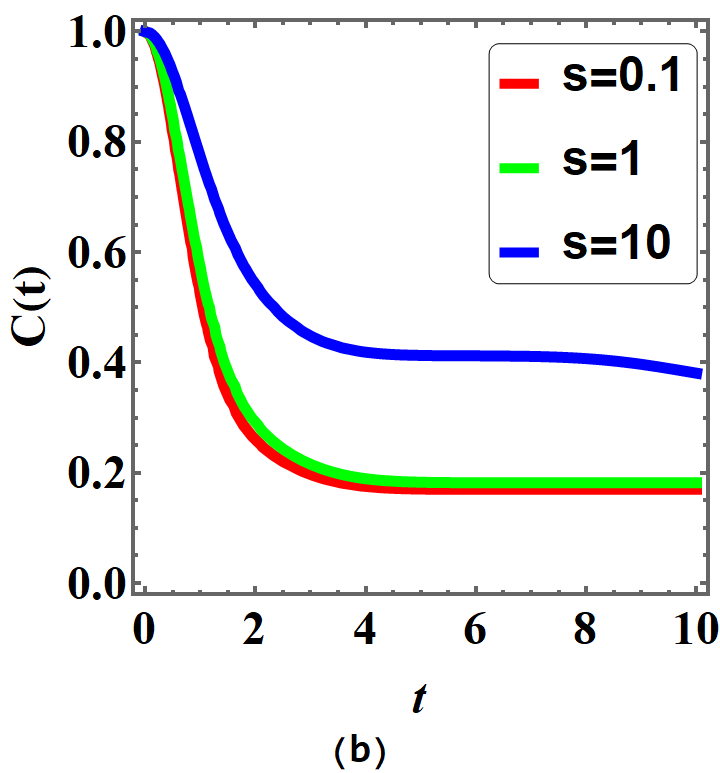}
	\caption{(a) Time variation of entanglement $C(t)$  We see that for different  separation  lengths between qubits, entanglement saturates  over long time scales with different values for (a) $\beta \Omega >>1$ and (b) $\beta \Omega <<1$.}
	\label{fig2}
\end{figure}

In order to understand the time evolution of concurrence, we first observe the behaviour of $\gamma(t)$ , $\gamma_0(t)$ and $\Phi(t)$ for different temperature regimes and time scales $s$. For this, we plot $\gamma(t)$  in figure \ref{fig1}(a) for low temperature  $\beta \Omega >>1$,  while  in figure \ref{fig1}(b) we have  $\gamma (t)$ is plotted for high temperature $\beta \Omega<<1$ case. From these plots, we observe that in the long time limit $\gamma(t)$ saturates at a certian value. We see that for short distances $L$ or small $s$, $\gamma(t)$ varies appreciably at high and low temperatures in comparison to  large $s$ values i.e. large L, Thus $L$ plays an important role to protect system from decoherence. Furthermore, from the non-monotonic  variation of $\gamma(t)$  leads to the identification of three regions of dynamics. First, for very short times $\gamma (t) \propto t^2$ showing strong non-Markovian behaviour while for the intermediate decay $\gamma (t) \propto t$ and thus Markovian behviour sets in the dynamics. Thirdly, the long time limit, there is a saturation region that implies  existence of finite value of quantum correlations. In figure \ref{fig1}(c), we have decay function due to initial correlations $\gamma_0(t)$, plotted for different values of $s$ and $\beta \Omega<<1$. We see that for small $s$, the $\gamma_0(t)$ has prounounced non-Markovian decay  in comparison to large $s$. Since concurrence has an  $e^{-\gamma_0(t)}$ factor, which therefore implies  initial system-bath correlations contribute to concurrence $C(t)$ for small $L$, while for large $L$, thermal fluctuations influence the decay. 
 Also, in fig \ref{fig1}(d), we plot $\Phi(t)$, for different values of $s$. we see that with the overall behvaiour of $\Phi(t)$ is same for all values of $s$, in other words, distance of separation of qubits doesnot  influence the dephasing. In the long time limit, $\Phi(t)$ goes to zero which implies that influnce of initial correlations on the entanglement remains over long periods of time.

Now, we plot time evolution of concurrence for different values of $s$ and temperature. At low temperatures $\beta \Omega >>1$, we see from the equations \ref{con}, \ref{gamma2} and \ref{phi2} that initial correlations $\gamma_0(t)$ do not contribute to entanglement dynamics. This can be understood as follows: at low temperatures, we have joint state of the system and bath $\rho= e^{-\beta H}/Z \to |GS\rangle \langle GS|$, where $|GS \rangle $ is some ground state of the Hamiltonian system (S) plus bath (B) $H_S+H_B$. For the model under consideration, $|GS\rangle = |11\rangle_{S} \otimes |0\rangle_B$. Thus $\rho_B^{\psi} = \frac{\bra{\psi}\exp(-\beta H)\ket{\psi}}{Tr_{B}\bra{\psi}\exp(-\beta H)\ket{\psi}}\to |0\rangle \langle 0|_B$. Therefore, $\rho_B^{\psi}$ has no non-trivial dependence on the system parameters  and thus we have an initally uncorrelated state.
 However, at high temperatures $\beta \Omega <<1$, we have $\gamma_0(t)$ influencing the concurrence. This can be seen from the plot for $\gamma_0(t)$ in figure \ref{fig1}(c). We see from this plot for small qubit separations, the correlations are least influenced by thermal fluctuations of the bath than the qubits with larger separation. The overall  contribution to concurrence from initial system-bath  correlations and damping factors $\gamma(t)$ is plotted in figures \ref{fig2} (a) and  (b) . In fig \ref{fig2}(a), we have concurrence plotted with respect $t$ at low temperatur $\beta \Omega >>1$. From this plot, we see that concurence decays in a non-Markovian way with saturation to a non-zero value of entaglement in the long time limit. Furthermore, concurrence variation at the high temperature case ($\beta \Omega <<1$) ( figure \ref{fig2}(b)) is mostly similiar in behaviour to low temperature case with  slightly lower vlaue for long time entanglement. We see from these plots, that qubit separation plays an important role in long time entangelemnt. The parameter can used to protect the entaglement decay in an optimal way. We observe from figures \ref{fig2}(a) and (b) that initial correlation superimposed with thermal fluctuations enhance the long time entanglement for large $s$ values in comparison to small $s$ or shorter qubits distances $L$. This can be intuitively understood in the following way. Since we are assuming a common bath coupled to two qubits which are  separated by certian distance. Therefore, there will be less energetic modes to scatter the qubits that would cause decoherence. In other words, at large $L$, there will be only few modes that couple to the qubits and lead to low  decoherence  rate while for small qubit separations, the number of modes increase (as we are integrating over solid angle in equation \ref{gamma2}) that cause fast   decoherence. However, the initial correlations suppress this rate so that we have finite concurrence in the long time limit. 

\section{Conclusions}
In conclusion, we have studied quantum entanglement between two qubits  coupled via  distance dependent interaction with a common bath. The initial correlated state is obtained via a projective measurement on the system while assuming a joint thermal equillibrium   state  of system and bath. Such procedure is important towards quantum state preparation where bath state can depend on the system parameters. Assuming, such initial system-bath  correlations, we studied their influence, and  distance of separation between the qubits on  dynamics of concurence and coherence in a large class of two qubit states. It is shown that  concurrence is half  of the coherence for all such kind of states, thus enabling to measure entanglement in terms of coherence. 

Next, we have shown that at low temperatures, initial system-bath correlations play no role in dynamics as the system and bath become uncorrelated. While at high temperatures, these correlations substantially modify the entanglement decay. Furthermore, we have shown  distance between the qubits forms an important parameter to control decoherece effects both at low and high temperature, i,e. we can tune $L$ in such a way that there exist only few modes which cause decoherence and some modes do not interact with qubits.   In order to characterize these modes and their influence in case of interacting qubits in equillbrium as well as non-equillibrium scenarios will be treated separately\cite{76}.

\begin{acknowledgements}
	The authors would like to thank Dr. Javid A Naikoo at University of Warsaw for helpful discussions.
\end{acknowledgements}
\appendix

\section*{Appendix}
\subsection{Bath Density Matrix $\rho_B^{\psi}(0)$}
In this appendix, we derive the state dependent bath density matrix $\rho_B^{\psi}(0)$. Since we assume a two qubit state 
$|\psi\rangle= a|00\rangle + b|01\rangle + c|10\rangle + d|11\rangle$ with $|a|^2 + |b|^2+|c|^2+|d|^2=1$ and bath density matrix is
\begin{align}
	\label{rhob}
	\rho_{B}^\psi = \frac{\bra{\psi}\exp(-\beta H)\ket{\psi}}{Tr_{B}\bra{\psi}\exp(-\beta H)\ket{\psi}}.
\end{align}
Using $ \sigma^z\ket{o}=\ket{0}, \sigma^z\ket{1}=-\ket{1}$, we have
\begin{eqnarray}
&&		e^{-\beta H}\ket{00} =  e^{-\beta \omega_0} e^{-\beta H_{B1}^+ }\ket{00},~~~
		e^{-\beta H}\ket{01} =   e^{-\beta H_{B1}^-}\ket{01}\\
&&	e^{-\beta H}\ket{10} =   e^{-\beta H_{B2}^+}\ket{10}~~~
		e^{-\beta H}\ket{11} =  e^{\beta \omega_0} e^{-\beta H_{B2}^-}\ket{11}
\end{eqnarray}
where 
\begin{eqnarray}
	H_{B1}^{\pm} = H_B + (B_{1k} \pm B_{2k}) ~~~H_{B2}^{\pm}= H_B- (B_{1k}\pm B_{2k})
\end{eqnarray}
and $B_{ik}= g_{ik}b_k+ g_{ik}^{\star}b_k^{\dagger}$. Therefore, we write
\begin{eqnarray}
	\label{part1}
	\langle \psi|e^{-\beta H} |\psi \rangle= |a|^2 e^{-\beta\omega_0}e^{-\beta H_{B1}^+} +|b|^2 e^{-\beta H_{B1}^-}+|c|^2 e^{-\beta H_{B2}^+}+ |d|^2 e^{\beta\omega_0} e^{-\beta H_{B2}^-}.
\end{eqnarray}
Next, the partition function $Z$, after a straight forward calculation, we write
\begin{eqnarray}
	\label{part}
	Z=Tr_B 	\langle \psi|e^{-\beta H} |\psi \rangle=
	\big[(|a|^2 e^{-\beta\omega_0} +|d|^2 e^{\beta\omega_0})e^{\beta \frac{\Sigma_{k}|g_{1k}+ g_{2k}|^2}{\omega_{k}}} +(|b|^2+|c|^2)e^{\beta \frac{\Sigma_{k}|g_{1k}- g_{2k}|^2}{\omega_{k}}}\big] Z_B \equiv z^{\prime} Z_B.
\end{eqnarray}
where $Z_B= Tr_B e^{-\beta H_B}$ and $z^{\prime }$ is the rest of the expression. Therefore, using \ref{part1} and \ref{part} in equation \ref{rhob}, we get the  bath density matrix depending on the system parameters.

\subsection{Time evolved density matrix $\rho_s(t)$}
In this appendix, we give explicit derivation of the time evolved density matrix given in equation \ref{densitymatrix}.
Suppose at time $t=0$ the state of the composite system is described by the initial density matrix $\rho(0)$,
then at time $t$ the density matrix in interaction picture is given by
\begin{eqnarray}
	\rho(t) = U(t)\rho(0)U(t)^{\dagger}
\end{eqnarray}
where $U(t)=T~e^{-i\int_0^t dt^{\prime} H_I(t^{\prime})}$ is the time evolution operator and $H_I(t)$ is the interaction
Hamiltonian in the interaction picture. Our main interest is to calculate the reduced density matrix of the system by tracing out degrees of freedom of the bath:
\begin{eqnarray}
	\rho_S(t)= {\rm Tr}_B[  U(t)\rho(0)U(t)^{\dagger} ].
\end{eqnarray}
We write 
$U(t) = e^{i\phi(t)}e^{\Sigma_{i}\sigma_{i}^{z}\hat\Lambda_i(t)}$ where $\phi(t)$
is a function of time only and $\hat\Lambda_i(t) = \Sigma_{k}(\alpha_{ik}(t)b_k - \alpha_{ik}^{*}(t)b_k^{\dagger})$ with $\alpha_{ik}(t) = g_{ik}\frac{(1-e^{-i\omega_{k}t})}{\omega_{k}}$ and $\phi(t)$ is a function of time only. Therefore, we can write
\begin{eqnarray}
	\rho_S(t) &=& {\rm Tr}_B[U(t) \rho(0) U(t)^{\dagger}]\\
	&=& {\rm Tr}_B[ e^{\Sigma_{i}\sigma_{i}^{z}\hat\Lambda_i(t)} |\psi\rangle \langle \psi| \otimes 
	\rho_B^{\psi}  e^{-\Sigma_{i}\sigma_{i}^{z}\hat\Lambda_i(t)}  ].
\end{eqnarray}
 	Let $\alpha_{1k}(t)+\alpha_{2k}(t)= A_k$ , $  \alpha_{1k}(t)-\alpha_{2k}(t)= B_k$ and
$\alpha_{1k}^*(t) + \alpha_{2k}^*(t)= A_k^*$,
$ \alpha_{1k}^*(t)-\alpha_{2k}^*(t) = B_k^*$
such that 
\begin{eqnarray}
	\label{6}
		&U(t)\ket{00} =  e^{i\phi(t)} \exp\Sigma_{k}( A_k^* b^\dagger -A_k b_k)\ket{00}, 
		~~~U(t)\ket{01} = e^{i\phi(t)}\exp\Sigma_{k}(B_k^* b^\dagger -B_k b_k)\ket{01},
		\\& U(t)\ket{10} = e^{i\phi(t)}\exp\Sigma_{k}(B_k b_k -B_k^* b_k^\dagger)\ket{10},
		~~~U(t)\ket{11} =  e^{i\phi(t)} \exp\Sigma_{k}(A_k b_k - A_k^* b_k^\dagger)\ket{11}.
\end{eqnarray}
We see that diagonal terms do not change, therefore we look for off-diagonal terms.
\begin{enumerate}
	\item $\boxed{\ket{00}\bra{01}}$ matrix element ($ab^*$):

	\begin{equation}\label{8}
		\begin{split}
		\varphi=	&\bra{00} \frac{1}{Z} \operatorname{Tr}_B \Big\{e^{\Sigma_{i}\sigma_{i}^{z}\hat\Lambda_i(t)}\ket{00}\bra{01}\otimes\Big[|a|^2 e^{-\beta\omega_0}e^{-\beta H_{B1}^+} +|b|^2 e^{-\beta H_{B1}^-}+|c|^2 e^{-\beta H_{B2}^+}+ |d|^2 e^{\beta\omega_0} e^{-\beta H_{B2}^-}\Big]e^{-\Sigma_{i}\sigma_{i}^{z}\hat\Lambda_i(t)}\Big\} \ket{01}\\
			& =\frac{1}{Z} \operatorname{Tr}_B \Big\{e^{\Sigma_{k}(A_{k}^{*}b_k^{\dagger} -A_{k}b_k)}\Big[ |a|^2 e^{-\beta\omega_0}e^{-\beta H_{B1}^+} +|b|^2 e^{-\beta H_{B1}^-}+|c|^2 e^{-\beta H_{B2}^+}+ |d|^2 e^{\beta\omega_0} e^{-\beta H_{B2}^-}\big]\Big\}.
		\end{split}
\end{equation}
Next we define unitary transformations
\begin{eqnarray}\label{9}
	O_{\pm} &=& \exp\mp[\Sigma_{k}( \frac{g_{1k}+g_{2k}}{\omega_{k}})b_k - (\frac{g^{*}_{1k}+g^{*}_{2k}}{\omega_{k}}b^\dagger_k)] \\
	\Theta_{\pm} &=& \exp\mp[\Sigma_{k}( \frac{g_{1k}-g_{2k}}{\omega_{k}})b_k - (\frac{g^{*}_{1k}-g^{*}_{2k}}{\omega_{k}}b^\dagger_k)]
\end{eqnarray}
such that
\begin{eqnarray}
&&	O_+ e^{-\beta H_{B1}^+} O_+^{\dagger}= e^{-\beta H_B} e^{\beta \frac{\Sigma_{k}|g_{1k}+ g_{2k}|^2}{\omega_{k}}}~~~~O_- e^{-\beta H_{B2}^+} O_-^{-1}= e^{-\beta H_B} e^{\beta \frac{\Sigma_{k}|g_{1k}+ g_{2k}|^2}{\omega_{k}}} \\
&& \Theta_{+} e^{-\beta H_{B1}^- }\Theta_{+}^{-1} = e^{-\beta H_B} e^{\beta \frac{\Sigma_{k}|g_{1k}- g_{2k}|^2}{\omega_{k}}}~~~\Theta_{-} e^{-\beta H_{B2}^- }\Theta_{-}^{-1} = e^{-\beta H_B} e^{\beta \frac{\Sigma_{k}|g_{1k}- g_{2k}|^2}{\omega_{k}}}.
\end{eqnarray}
Next, using these unitary transformations, we write
\begin{eqnarray}
&&	{\rm Tr}_B \big[ e^{\Sigma_{k}(A_{k}^{*}b_k^{\dagger} -A_{k}b_k)} e^{-\beta H_{B1}^+} e^{\Sigma_{k}(B_{k}b_k -B_k^*b_k^\dagger)} \nonumber \\
&&~~~~~= {\rm Tr}_B \big[ e^{\Sigma_{k}(A_{k}^{*}b_k^{\dagger} -A_{k}b_k)} O_+^{-1} O_+ e^{-\beta H_{B1}^+} O_+^{-1} O_+e^{\Sigma_{k}(B_{k}b_k -B_k^*b_k^\dagger)} \big] \nonumber \\
&&~~~~~= e^{\beta \Sigma_{k}\frac{|g_{1k}+ g_{2k}|^2}{\omega_{k}}}\operatorname{Tr}_B \Big\{e^{A_k^*b_k^\dagger}e^{-A_kb_k} e^{-\frac{1}{2} |A_k|^2}O_+^{-1}e^{-\beta H_B}O_+ e^{B_k b_k}e^{-B_k^* b_k^\dagger}e^{\frac{1}{2} |B_k|^2}\Big\} \nonumber \\
&&~~~~~=e^{\beta \Sigma_{k}\frac{|g_{1k}+ g_{2k}|^2}{\omega_{k}}}\operatorname{Tr}_B \Big\{ O_+^{-1} O_+ e^{A_k^*b_k^\dagger} O_+^{-1} O_+ e^{-A_kb_k} e^{-\frac{1}{2} |A_k|^2}O_+^{-1}e^{-\beta H_B}O_+ e^{B_k b_k} O_+^{-1} O_+e^{-B_k^* b_k^\dagger}e^{\frac{1}{2} |B_k|^2}\Big\} \nonumber \\
&&~~~~=e^{\beta \Sigma_{k} \frac{|g_{1k}+ g_{2k}|^2}{\omega_{k}}}e^{\Sigma_{k} \frac{(g_{1k}+g_{2k})(B_k^*-A_k^*)}{\omega_{k}}}
 e^{\Sigma_{k}\frac{-(g_{1k}^*+g_{2k}^*)(B_k-A_k)}{\omega_{k}} }
 e^{\Sigma_{k} \frac{(A_k^*B_k-B_k^* A_k)}{2} } e^{-\gamma_1(t)}
\end{eqnarray}
where $e^{-\gamma_1(t)}= \operatorname{Tr}_B\big\{e^{(B_k-A_k)b_k-(B_k^*-A_k^*)b_k^\dagger}\rho_{B}\big\}$. In the similiar fashion, we can evluate other terms. Thus putting together all these terms, 
 and using the form of  $Z$ in equation \ref{8}, we arrive at the following expression for the matrix element $|00\rangle \langle 01|$ with coefficient $a^*b \varphi(t)$ where 
\begin{equation}\label{17} 
		\begin{split}
	\varphi(t)=	&\frac{e^{-\gamma_1(t)}}{z'}\Bigg\{|a|^2 e^{-\beta\omega_0}e^{\beta \Sigma_{k}\frac{|g_{1k}+ g_{2k}|^2}{\omega_{k}}}\exp \Sigma_{k} {\frac{4i|g_k|^2}{\omega_{k}^2}[1+\cos k.(r_1-r_2)\sin \omega_{k}t]}\\&+|b|^2e^{\beta \Sigma_{k}\frac{|g_{1k}- g_{2k}|^2}{\omega_{k}}}
			\exp {- \Sigma_{k}\frac{4i|g_k|^2}{\omega_{k}^2}[1-\cos k.(r_1-r_2)\sin \omega_{k}t]}\\&+|c|^2e^{\beta \Sigma_{k}\frac{|g_{1k}- g_{2k}|^2}{\omega_{k}}}\exp {\Sigma_{k}\frac{4i|g_k|^2}{\omega_{k}^2}[2\sin k.(r_1-r_2)(1-\cos \omega_{k}t)+\sin \omega_{k}t(1-\cos k.(r_1-r_2))]}\\&+|d|^2 e^{\beta\omega_0}e^{\beta \Sigma_{k}\frac{|g_{1k}+ g_{2k}|^2}{\omega_{k}}}\exp \Sigma_{k} {\frac{4i|g_k|^2}{\omega_{k}^2}[2\sin k.(r_1-r_2)(1-\cos \omega_{k}t)-\sin \omega_{k}t(1+\cos k.(r_1-r_2))]}\Bigg\},
	\end{split}
\end{equation}
and  after further simplication, we arrive at the final expression 
\begin{eqnarray}
	\varphi(t) &=& \Bigg[\big\{|a|^2 e^{-\beta\omega_0} e^{i\Phi_+(r,t)}+|d|^2 e^{\beta\omega_0}e^{i(\chi(r,t)-\Phi_+(r,t))}\big\}e^{\beta\Sigma_k\frac{|g_{1k}+g_{2k}|^2}{\omega_k}}  \nonumber \\
	&&
	+\big\{|b|^2 e^{-i\Phi_-(r,t)} +|c|^2 e^{i(\chi(r,t)+\Phi_-(r,t))}\big\}e^{\beta\Sigma_k\frac{|g_{1k}-g_{2k}|^2}{\omega_k}}\Bigg] \exp(-4\Sigma_{k}\frac{|g_k|^2}{\omega_{k}^2}(1-\cos{\omega_{k}t}) \coth{\frac{\beta\omega_{k}}{2}}) 
\end{eqnarray}
where 
\begin{eqnarray}
	\label{phis}
	\Phi_\pm(r,t)&=&4\sum_k\frac{|g_k|^2}{\omega_k^2}\Big[\sin{\omega_k t}(1\pm\cos \vec{k}.(\vec{r}_1-\vec{r}_2)) \Big]\\
	\chi(r,t)&=&4\sum_k \frac{|g_k|^2}{\omega_k^2}\big[\sin \vec{k}.(\vec{r}_1-\vec{r}_2)(1-\cos \omega_kt)\big]
\end{eqnarray}

In the similiar way, we obtain all other matrix elements which are given by:
	
\item $|00\rangle \langle 10|$ matrix element ($ac^*$):	
\begin{eqnarray}
\zeta(t) &=	&\bra{00} \frac{1}{Z} \operatorname{Tr}_B \Big\{e^{\Sigma_{i}\sigma_{i}^{z}\hat\Lambda_i(t)}\ket{00}\bra{10}\otimes\Big[|a|^2 e^{-\beta\omega_0}e^{-\beta H_{B1}^+} +|b|^2 e^{-\beta H_{B1}^-}+|c|^2 e^{-\beta H_{B2}^+}+ |d|^2 e^{\beta\omega_0} e^{-\beta H_{B2}^-}\Big]e^{-\Sigma_{i}\sigma_{i}^{z}\hat\Lambda_i(t)}\Big\} \ket{10} \nonumber \\
	& =& \Bigg[\big\{|a|^2 e^{-\beta\omega_0} e^{i\Phi_+(r,t)}+|d|^2 e^{\beta\omega_0} e^{-i{(\chi(r,t)-\Phi_+(r,t))}}\big\}e^{\beta\Sigma_k\frac{|g_{1k}+g_{2k}|^2}{\omega_k}} \nonumber\\
&&~+\big\{|b|^2 e^{-{i(\chi(r,t)-\Phi_-(r,t))}} +|c|^2 e^{-i\Phi_-(r,t)}\big\}e^{\beta\Sigma_k\frac{|g_{1k}-g_{2k}|^2}{\omega_k}}\Bigg]\exp(-4\Sigma_{k}\frac{|g_k|^2}{\omega_{k}^2}(1-\cos{\omega_{k}t}) \coth{\frac{\beta\omega_{k}}{2}})
\end{eqnarray}

\item $|00\rangle \langle 11|$ matrix element ($ad^*$):

\begin{eqnarray}
\kappa(t)	&=	&\bra{00} \frac{1}{Z} \operatorname{Tr}_B \Big\{e^{\Sigma_{i}\sigma_{i}^{z}\hat\Lambda_i(t)}\ket{00}\bra{11}\otimes\Big[|a|^2 e^{-\beta\omega_0}e^{-\beta H_{B1}^+} +|b|^2 e^{-\beta H_{B1}^-}+|c|^2 e^{-\beta H_{B2}^+}+ |d|^2 e^{\beta\omega_0} e^{-\beta H_{B2}^-}\Big]e^{-\Sigma_{i}\sigma_{i}^{z}\hat\Lambda_i(t)}\Big\} \ket{11} \nonumber \\
	& =& \Bigg[\big\{|a|^2 e^{-\beta\omega_0} e^{2i\Phi_+(r,t)}+|d|^2 e^{\beta\omega_0} e^{-2i{(\Phi_+(r,t)}}\big\}e^{\beta\Sigma_k\frac{|g_{1k}+g_{2k}|^2}{\omega_k}} \nonumber \\
	&&~+\big\{|b|^2 e^{-{i(\chi(r,t)}} +|c|^2 e^{i\chi(r,t)}\big\}e^{\beta\Sigma_k\frac{|g_{1k}-g_{2k}|^2}{\omega_k}}\Bigg]\exp(-4\Sigma_{k}\frac{|g_k|^2}{\omega_{k}^2}\big(1+\cos{k.(r_1-r_2)}\big)(1-\cos{\omega_{k}t}) \coth{\frac{\beta\omega_{k}}{2}})
\end{eqnarray}
	
\item $ |01\rangle \langle 10|$ matrix element ($bc^*$):

\begin{eqnarray}
	\bar{\kappa}(t) 	&=	&\bra{01} \frac{1}{Z} \operatorname{Tr}_B \Big\{e^{\Sigma_{i}\sigma_{i}^{z}\hat\Lambda_i(t)}\ket{01}\bra{10}\otimes\Big[|a|^2 e^{-\beta\omega_0}e^{-\beta H_{B1}^+} +|b|^2 e^{-\beta H_{B1}^-}+|c|^2 e^{-\beta H_{B2}^+}+ |d|^2 e^{\beta\omega_0} e^{-\beta H_{B2}^-}\Big]e^{-\Sigma_{i}\sigma_{i}^{z}\hat\Lambda_i(t)}\Big\} \ket{10} \nonumber \\
		&=& \Bigg[\big\{|a|^2 e^{-\beta\omega_0} e^{i\chi(r,t)}+|d|^2 e^{\beta\omega_0} e^{-i(\chi(r,t)}\big\}e^{\beta\Sigma_k\frac{|g_{1k}+g_{2k}|^2}{\omega_k}} \nonumber \\
		&&~+\big\{|b|^2 e^{2{i\Phi_-(r,t)}} +|c|^2 e^{-2i\Phi_-(r,t)}\big\}e^{\beta\Sigma_k\frac{|g_{1k}-g_{2k}|^2}{\omega_k}}\Bigg]\exp(-4\Sigma_{k}\frac{|g_k|^2}{\omega_{k}^2}\big(1-\cos{k.(r_1-r_2)}\big)(1-\cos{\omega_{k}t}) \coth{\frac{\beta\omega_{k}}{2}})
\end{eqnarray}
\item $ |01\rangle \langle 11|$ matrix element ($bd^*$):
\begin{eqnarray}
	\bar{\zeta}(t)	&=	&\bra{01} \frac{1}{Z} \operatorname{Tr}_B \Big\{e^{\Sigma_{i}\sigma_{i}^{z}\hat\Lambda_i(t)}\ket{01}\bra{11}\otimes\Big[|a|^2 e^{-\beta\omega_0}e^{-\beta H_{B1}^+} +|b|^2 e^{-\beta H_{B1}^-}+|c|^2 e^{-\beta H_{B2}^+}+ |d|^2 e^{\beta\omega_0} e^{-\beta H_{B2}^-}\Big]e^{-\Sigma_{i}\sigma_{i}^{z}\hat\Lambda_i(t)}\Big\} \ket{11} \nonumber \\
& = &\Bigg[\big\{|a|^2 e^{-\beta\omega_0} e^{i(\chi(r,t)+\Phi_-(r,t))}+|d|^2 e^{\beta\omega_0} e^{-i{\Phi_+(r,t)}}\big\}e^{\beta\Sigma_k\frac{|g_{1k}+g_{2k}|^2}{\omega_k}} \nonumber\\
&&~ +\big\{|b|^2 e^{{i\Phi_-(r,t)}} +|c|^2 e^{i(\chi(r,t)-\Phi_-(r,t))}\big\}e^{\beta\Sigma_k\frac{|g_{1k}-g_{2k}|^2}{\omega_k}}\Bigg]\exp(-4\Sigma_{k}\frac{|g_k|^2}{\omega_{k}^2}(1-\cos{\omega_{k}t}) \coth{\frac{\beta\omega_{k}}{2}})
\end{eqnarray}

\item $ |10\rangle \langle 11|$ matrix element ($bc^*$):

\begin{eqnarray}
\bar{\varphi}(t) 		&=	&\bra{10} \frac{1}{Z} \operatorname{Tr}_B \Big\{e^{\Sigma_{i}\sigma_{i}^{z}\hat\Lambda_i(t)}\ket{10}\bra{11}\otimes\Big[|a|^2 e^{-\beta\omega_0}e^{-\beta H_{B1}^+} +|b|^2 e^{-\beta H_{B1}^-}+|c|^2 e^{-\beta H_{B2}^+}+ |d|^2 e^{\beta\omega_0} e^{-\beta H_{B2}^-}\Big]e^{-\Sigma_{i}\sigma_{i}^{z}\hat\Lambda_i(t)}\Big\} \ket{11} \nonumber \\
	&=& \Bigg[\big\{|a|^2 e^{-\beta\omega_0} e^{-i(\chi(r,t)-\Phi_+(r,t))}+|d|^2 e^{\beta\omega_0} e^{-i{\Phi_+(r,t)}}\big\}e^{\beta\Sigma_k\frac{|g_{1k}+g_{2k}|^2}{\omega_k}} \nonumber\\
	&&~+\big\{|b|^2 e^{-{i(\chi(r,t)+\Phi_-(r,t))}} +|c|^2 e^{i\Phi_-(r,t)}\big\}e^{\beta\Sigma_k\frac{|g_{1k}-g_{2k}|^2}{\omega_k}}\Bigg]\exp(-4\Sigma_{k}\frac{|g_k|^2}{\omega_{k}^2}(1-\cos{\omega_{k}t}) \coth{\frac{\beta\omega_{k}}{2}})
\end{eqnarray}
	
\end{enumerate}
Unisng all these equations, we get the time evolved density matrix used in the main text \ref{densitymatrix}:
\begin{eqnarray}\label{43}
	\rho_{s}(t)=\begin{pmatrix}
		|a|^2 & ab^* \varphi(t) & ac^*\zeta(t)& ad^* \kappa(t)\\
		ba^* \varphi^*(t) & |b|^2 & bc^*\bar{\kappa}(t) &bd^*\bar{\zeta}(t)\\
		ca^*\zeta^*(t)&cb^*\bar{\kappa}^*(t)&|c|^2&cd^*\bar{\varphi}(t)\\
		da^* \kappa^*(t)&db^*\bar{\zeta}^*(t)&dc^* \bar{\varphi}^*(t)& |d|^2
	\end{pmatrix}
\end{eqnarray}

\end{document}